\documentclass[3p,times,twocolumn]{elsarticle}
 \biboptions{comma,sort&compress}

\usepackage{graphicx}
\usepackage{amsmath}
\usepackage{here}
\usepackage{ecrc}

\volume{00}

\firstpage{1}

\journalname{Nuclear and Particle Physics Proceedings}
\runauth{Carlos Quezada-Calonge}

\jid{nppp}
\jnltitlelogo{Nuclear and Particle Physics Proceedings}

\usepackage{amssymb}
\usepackage{amsmath,amsthm}

\usepackage[figuresright]{rotating}

\usepackage{amsmath,amsthm}

\usepackage{slashed}
\usepackage{amssymb}

\usepackage{calrsfs}
\usepackage{xcolor}
\usepackage{graphicx}

\usepackage{float}
\usepackage{subcaption} 

\usepackage{physics}

\usepackage{comment} 
\usepackage{tabularx}

\usepackage{booktabs}

\usepackage{savesym}

\savesymbol{darray}

\usepackage{nccmath}

\usepackage{color}

\newcommand{\mL}{\mathcal{L}}

\newcommand{\mO}{\mathcal{O}}

\renewcommand\[{\begin{dmath}}
\renewcommand\]{\end{dmath}}

\begin{document}

\begin{frontmatter}

\title{
%
How relevant are top loops in VBS at the LHC?}

\author{Carlos Quezada-Calonge\corref{cor0} }  
\ead{cquezada@ucm.es}

\author{Antonio Dobado}
\ead{dobado@fis.ucm.es}

\author{Juan Jos\'e Sanz-Cillero}
\ead{jjsanzcillero@ucm.es}

\cortext[cor0]{\vspace*{-0.cm}Speaker, corresponding author.
} 
\cortext[cor1]{Talk given at 25th International Conference in Quantum Chromodynamics (QCD 22),  4 - 7 July 2022, Montpellier - FR. This research is partly supported by MICIN under grant PID2019-108655GB-I00/AEI/10.13039/501100011033, UCM under research group 910309, IPARCOS institute, and by EU STRONG-2020 under the program H2020-INFRAIA-2018-1 [grant no.824093].}
\address{Departamento de F\'{\i}sica Te\'{o}rica and IPARCOS,  \\ 
Universidad Complutense de Madrid, Plaza de las Ciencias 1, 28040-Madrid, Spain} 

\pagestyle{myheadings}
\markright{ }
\begin{abstract}
\noindent

We present the contributions to the imaginary part of the $W^+W^-$ elastic scattering from top and bottom quark one-loop diagrams. 
The computation is performed within the context of HEFT, where we compare them with the boson-loop corrections. We argue that the often neglected (top and bottom) fermion contributions may in fact be relevant.

\end{abstract}
\end{frontmatter}
\section{Introduction}

\vspace*{-0.3cm}In the absence of direct evidence of New Physics (NP) manifested as new 
particles in colliders, much effort has been put into the study of higher order corrections in order to elucidate the possible ultraviolet (UV) theory that lies underneath. One of the processes where this NP might show up is Vector Boson Scattering (VBS). This very absence of new states suggests the existence of a mass gap, a scale of NP far above our current reach. In this context, Effective Field Theories (EFT) are the right tool to probe the different NP scenarios. One of the most general EFT available is the so-called Higgs Effective Field Theory (HEFT). 
Higher order corrections to VBS are often studied using HEFT since bosonic-loops formally scale like $\mathcal{O}(s^2/v^4)$ and deviations from the Standard Model (SM) might point out towards a strongly interacting NP symmetry breaking sector. Additionally, we also have fermion-loop corrections which formally scale like $\mathcal{O}(M_{\rm Fer}^2 s/v^4)$. Though boson corrections might be dominant as they increase rapidly with the center-of-mass energy there is a point to be made in addressing how relevant fermionic corrections are when one considers the heaviest fermions ($t$ and $b$ quarks), the available HEFT coupling parameter space and the energy regime. 

These proceedings summarize the results found in our most recent study \cite{Quezada-Calonge:2022lop}. We use HEFT to  calculate the imaginary part of $t$ and $b$ loop corrections in $W^+ W^- \rightarrow W^+  W^-$ scattering and then compare them with the imaginary part arising from boson loops. We focus on the imaginary part since it first appears at Next to Leading Order (NLO) in the chiral expansion and cannot be masked by the purely real Lowest Order (LO) order amplitude. 
In this way, a large imaginary part from fermion loops would be an indication of similarly large fermion loop corrections in the real part, and therefore they should not be neglected with respect to boson loops. 

\section{HEFT Lagrangian}

Our Lagrangian will be given by the lowest order operators of chiral dimension $\mO(p^2/v^2)$~\cite{Appelquist,EWChL-HEFT},
\begin{eqnarray}
\mL_{2} &=&  \mL_S\,  +\,  \mathcal{L}_{\rm Yuk} \,+\, \mL_{\rm kin-F} \, \, +\, \mathcal{L}_{\rm YM}\, \, , 
\label{eq:HEFT-Lagr}
\end{eqnarray}
with
\vspace*{-0.89cm} 
\begin{eqnarray}
\phantom{ZZZZ} \mathcal{L}_S  &=&   \frac{v^4}{4} \mathcal{F}(h) {\rm Tr}\{( D_\mu U)^\dagger D^\mu U\} 
  +\frac{1}{2}  (\partial h)^2 
  - V(h) \,,
\nonumber \\
\mL_{\rm kin-F} &=& i \bar{t}  D \hspace*{-0.2cm} \slash\, t
\, + \,  i \bar{b}  D \hspace*{-0.2cm} \slash\, b \,,
\label{eq:Yuk-Lagr}
\end{eqnarray}
where the covariant derivatives in $\mL_{\rm kin-F}$ and $\mL_S$  contain the couplings with the EW gauge bosons and $\mathcal{L}_{\rm YM}$ is the standard $SU(2)_L\times U(1)_Y$ Yang-Mills Lagrangian. The Yukawa Lagrangian we are interested in is:
\begin{eqnarray}
\mathcal{L}_{\rm Yuk} & =&-\mathcal{G}(h) \bigg [\sqrt{ 1-\frac{ \omega^2  }{v^2}}(M_t \bar{t} t+M_b \bar{b}b)
\nonumber \\ 
&&  +i\frac{\omega ^0}{v}\left( M_t \bar{t}\gamma^5t-M_b\bar{b}\gamma^5b \right)  
\nonumber \\ 
&&  + \, 
\left[ i\frac{ \sqrt{2} \omega ^+}{v}\left(M_b \bar{t} P_R b-M_t \bar{t} P_L b   \right) +\mbox{h.c.}\right] \, ,
\label{eq:Yuk-Lagr}
\end{eqnarray}
with the projectors $P_{R,L}=\frac{1}{2}(1\pm \gamma_5)$ and finally the Higgs functions are defined as:
\begin{eqnarray}
    \mathcal{G}(h)&=&1+{c_1} \frac{h}{v}+{c_2} \frac{h^2}{v^2}+{c_3}\frac{h^3}{v^3}+{c_4}\frac{h^4}{v^4}+... \, ,
    \nonumber\\
    \mathcal{F}(h)&=&1+2{a} \frac{h}{v}+{b} \frac{h^2}{v^2}+{c} \frac{h^3}{v^3}+...\, ,
    \nonumber\\
    V(h)&=& \frac{M_h^2}{2}  h^2+ d_3 \frac{M_h^2}{2v^2}  h^3+ {d_4} \frac{M_h^2}{8v^2} h^4+...
\end{eqnarray}
In the SM case one has $a=b=1$ and $c=0$, for $\mathcal{F}(h)$, $c_1=1$ and $c_{n \geq 2}=0$ in $\mathcal{G}(h)$, and $d_3=d_4=1$ and $d_{i\geq 5}$ in $V(h)$. 

\section{Fermion and boson loops}

In this section we will enunciate how the imaginary parts are calculated, we will introduce the partial waves and we will define the relevant quantities
to be computed.

As we commented above, we will be focusing on the imaginary part of top and bottom loops contributing to $W^+ W^- \rightarrow W^+  W^-$ and compare them with the imaginary part of bosonic loops of the same process. We do this since they first appear at NLO in the chiral expansion at $\mathcal{O}(p^4/v^4)$ and they are not masked by the purely real LO amplitude $\mathcal{O}(p^2/v^2)$. Formally this means  $\mbox{Im}\mathcal{A}=\mbox{Im}\mathcal{A}_{4,1\ell}$, where the amplitude can be expanded in terms of the following Partial Wave Amplitudes (PWAs):

\begin{equation} 
\mathcal{A}(s,t)=\sum_J 16\pi K (2J+1) P_J(\cos\theta) \, a_J(s)\, ,
\end{equation} 
with $K=1$ ($K=2$) for distinguishable (indistinguishable) final particles. In the physical energy region, Im~$a_J(s)$ will be provided by the one-loop absorptive cuts in the $s$-channel, which we will use to label the various contributions.

In order to calculate the imaginary part of the loops we will use  perturbative unitarity in terms of the tree level amplitudes. The collection of intermediate states (loops) of fermions and bosons is provided by, 
\begin{eqnarray}
 {\rm Fer}_J&=&\mbox{Im}\, a_J|_{b\bar{b}      ,t\bar{t}}    \, ,\nonumber\\ 
 {\rm Bos}_J&=&\mbox{Im}\ a_J|_{\gamma \gamma, \gamma Z, \gamma h,W^+ W^-,ZZ, Zh, hh}      \, .
 \label{eqn:cuts}
\end{eqnarray}

Additionally, it is important to keep in mind which HEFT couplings enter in the contribution to each PWA:
\begin{eqnarray}
\begin{split}
\mbox{\bf $J=0$:} \qquad 
&{\rm Fer}_0 \quad \longrightarrow\quad  a,c_1 ,\\
&{\rm Bos}_0 \hspace{-0.067cm} \quad \longrightarrow\quad a,b, d_3, \\
\mbox{\bf $J=1$:} \qquad 
&{\rm Fer}_1  \quad\longrightarrow\quad  \mbox{no \  dependence,  } \\ 
& {\rm Bos}_1 \hspace{-0.067cm} \quad\longrightarrow\quad  a\, .
\end{split}
\label{eq:heftdep}
\end{eqnarray}

The tree level amplitude $\mathcal{A}(W^+W^-\to F\overline{F})\equiv Q^{\Delta \lambda,F}$ (one for the production of each intermediate fermion state $F\overline{F}$), with  $Q^{0,\, F}=\frac{1}{\sqrt{2}}(Q^{++, F}-Q^{--,F})=\sqrt{2} Q^{++,F} $, $ Q^{+-,F}$  and $Q^{-+,F}$. For $J=0$ only the $Q^{0,F}$ combination is necessary for the partial-wave projection $Q_J^{\Delta \lambda,F}$, while for $J=1$ three  $\Tilde{Q}^F=$ $(Q^{0,F}, Q^{ +-,F}, Q^{-+,F})$ enter in the projection:
\begin{equation}
\hspace*{-0.6cm} Q^{\Delta \lambda }_J \, \hspace*{-0.05cm} = \hspace*{-0.05cm}\, \frac{1}{64\pi^2 K}\sqrt{\frac{4\pi}{2J+1}} \int  Q^{\Delta \lambda}(s,\Omega)\, Y^*_{J,\Delta \lambda}(\Omega)\, d\Omega\,,  \end{equation}

where $Y_{JM}(\Omega)$ are the spherical harmonics  and $\Delta \lambda$ is the helicity difference $\Delta \lambda=\lambda_1-\lambda_2$, with the super-index $F$ omitted for simplicity and $\beta_F=\sqrt{1-4 M_F^2/s}$. Hence the projected amplitudes are:

\begin{eqnarray}
     \rm{Fer}_0 &\hspace*{-0.05cm}=& \hspace*{-0.05cm} \mbox{Im}\  a_0(s)\bigg|_{t\bar{t},b\bar{b}}  \, \hspace*{-0.1cm} =\, \hspace*{-0.3cm} \displaystyle{\sum_{F=t,b}}   \beta_F \, \left|Q^{0,\, F}_{0}\right|^2 \,  \hspace*{-0.1cm} \theta(s-4M_F^2)   ,
\\
    \rm{Fer}_1 &\hspace*{-0.05cm}=& 
    \hspace*{-0.05cm} \mbox{Im} \ a_1(s)\bigg|_{t\bar{t},b\bar{b}}  \, \hspace*{-0.1cm}=\,\hspace*{-0.3cm} \displaystyle{\sum_{F=t,b}}  
    \beta_F\,  \left|\tilde{Q}^{n,F}_1\right|^2 \theta(s-4M_F^2)  , 
%
\end{eqnarray}
where $\tilde{Q}^{n,F}$ is the n-th element of $\tilde{Q}^{F}$ and a sum from 1 to 3 over $n$ is understood .

To study the fermion relevance we introduce the ratio: 
\begin{ceqn}
\begin{align}
    R_J=\frac{{\rm Fer}_J}{{\rm Bos}_J+{\rm Fer}_J} .
\end{align}
\end{ceqn}
Values of $R_J$ close to zero will indicate that we can safely drop fermion-loops, while deviations from this value will point out the relevance of fermions in $WW$ scattering. Additionally we can define the cumulative relative ratios of each cut as $\chi^J_i= \sum_{n=1}^{i} \mbox{Im}\ a_J\, \bigg|_{n} / \mbox{Im}\ a_J$  where $\rm N_{ch}$ is the total number of absorptive channels, and Im $ a_J= \displaystyle{ \sum_{n=1}^{\rm N_{ch}} \mbox{Im}\ a_J\bigg|_n }$ is the total imaginary part of the $a_J$~PWA. 

Finally, when projecting the one-loop amplitudes onto PWA's over the full angular domain  a singularity arises from photon exchanges in the t-channel for the $W^+W^-$ cut. In order to deal with this divergence we have followed two approaches: a) Assuming the $g'=0$ limit (custodial limit in the bosonic sector), which decouples the photon and hence the singularity disappears; and b) imposing an angular cut on all partial-wave projections, which avoids the singularity. The former option is often used in purely bosonic analyses. Given the experimental suppression of the ratio $g^{'2}/g^2 \approx (M_Z^2-M_W^2)/M_W^2\ll 1$ this approximation $g'=0$ is in general considered reasonable. The latter approach also solves the singularity problem but presents other issues, as these angular-cut pseudo PWA (p-PWA) are no longer orthogonal to each other. On the other hand, this option allows us to incorporate all physical absorptive cuts.  More comments on this can be found in~\cite{Quezada-Calonge:2022lop}.   

Concerning the center-of-mass energy we have studied the interval  $0.5$~TeV$\leq \sqrt{s}\leq 3$~TeV, which is the relevant one to look for NP at the LHC. We will use as inputs: $M_W$=80.38~GeV, $M_Z$=91.19~GeV, $M_H$=125.25~GeV, $v$=246.22~GeV, $M_t$=172.76~GeV and $M_b$= 4.18~GeV~\cite{pdg}. The value of the Weinberg angle is found in the standard way from $M_W$ and $M_Z$, $\cos^2 \theta_W = M_W^2/M_Z^2$ at LO. 
 For our analysis, we will allow a 10\% deviation with respect to the SM in the HEFT couplings. These variations are of the order of the uncertainties found experimentally for $a$ and $c_1$~\cite{deBlas:2018tjm}. Expecting their improvement in the future, we have considered similar variations for $b$ and $d_3$, even if  their current experimental errors are still $\mO(100\%)$~\cite{ATLAS:2020jgy}.  



{\bf $\bullet$ \underline{$R_0$ in the $g'=0$ limit:}}

\begin{figure}[!t]  
\begin{subfigure}{1 \columnwidth}
  \centering
  \includegraphics[width=0.85\columnwidth]{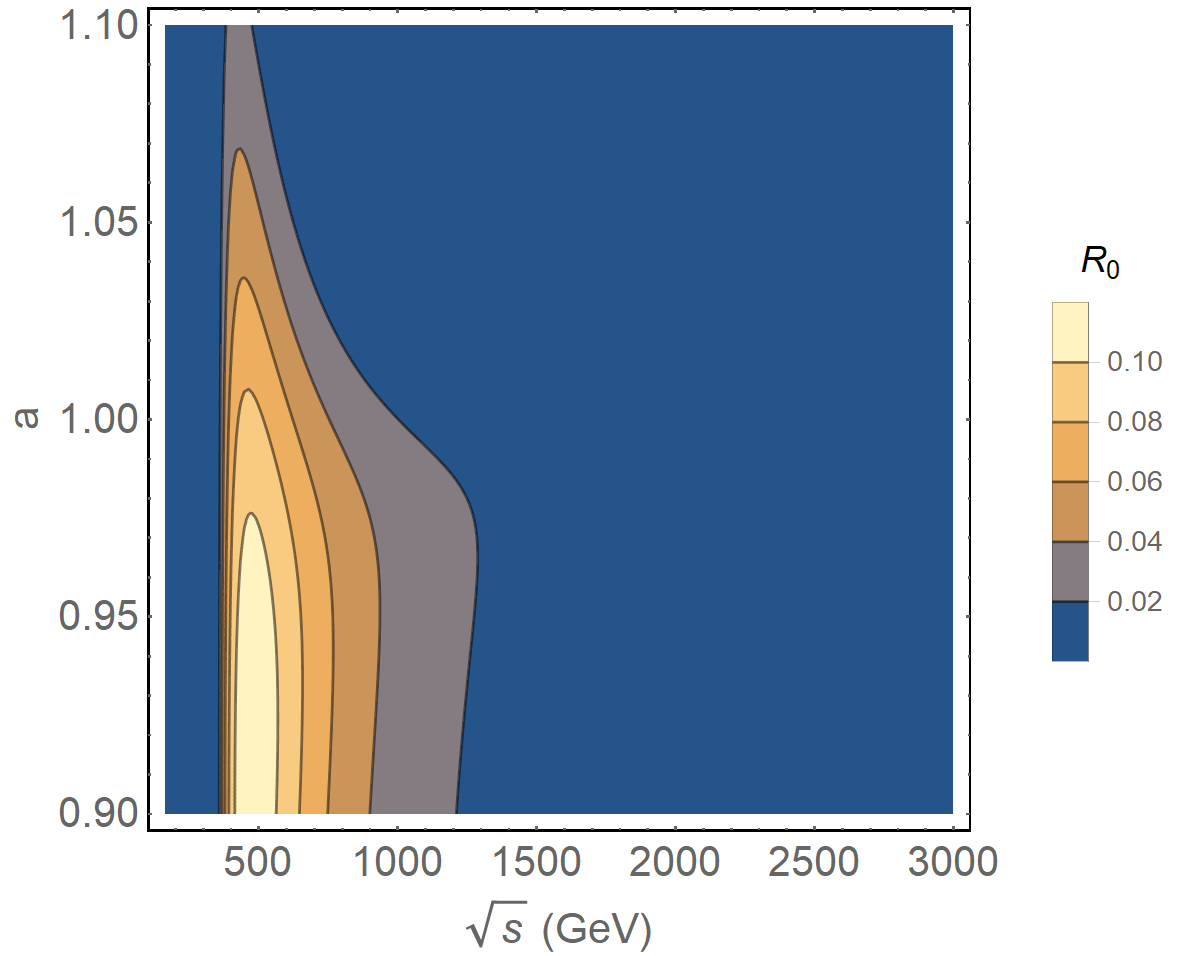}  
  \caption{$R_0$ dependence on $a$ for $b=c_1=d_3=1$.}
  \label{fig:r0_countour_a_isospin}
\end{subfigure}\\[0.5ex]

\begin{subfigure}{1\columnwidth}
  \centering
  \includegraphics[width=0.85\columnwidth]{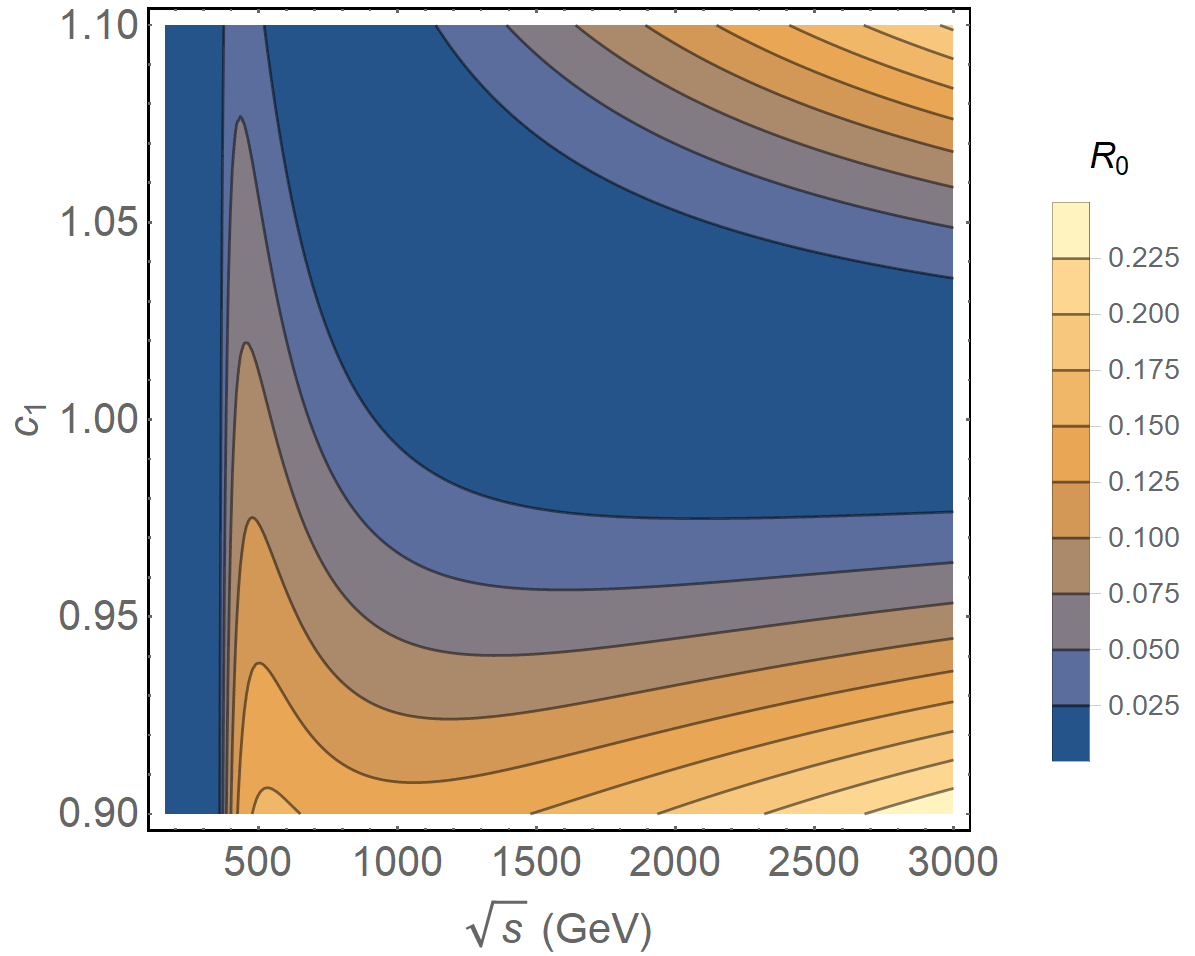}  
  \caption{$R_0$ dependence on $c_1$ for $a=b=d_3=1$.}
  \label{fig:r0_countour_c1_isospin}
\end{subfigure}

\caption{}
\label{fig:figr0contour}
\end{figure}

Fig.~\ref{fig:figr0contour} illustrates the situation of $R_0$ with respect to the modification of one coupling at a time. We can observe the dependence on $a$ in Fig.~\ref{fig:r0_countour_a_isospin}, where boson loops completely dominate at high energies. 
The dependence on $b$ is similar and the variation with $d_3$ is negligible, so they will not be shown here.  
On the other hand, Fig.~\ref{fig:r0_countour_c1_isospin} shows that we can find fermion corrections of roughly 22\% for the maximum deviation of $c_1$.

\begin{figure}[!t]    
\centering

\begin{subfigure}{1 \columnwidth}
  \centering
  \includegraphics[width=0.8 \columnwidth]{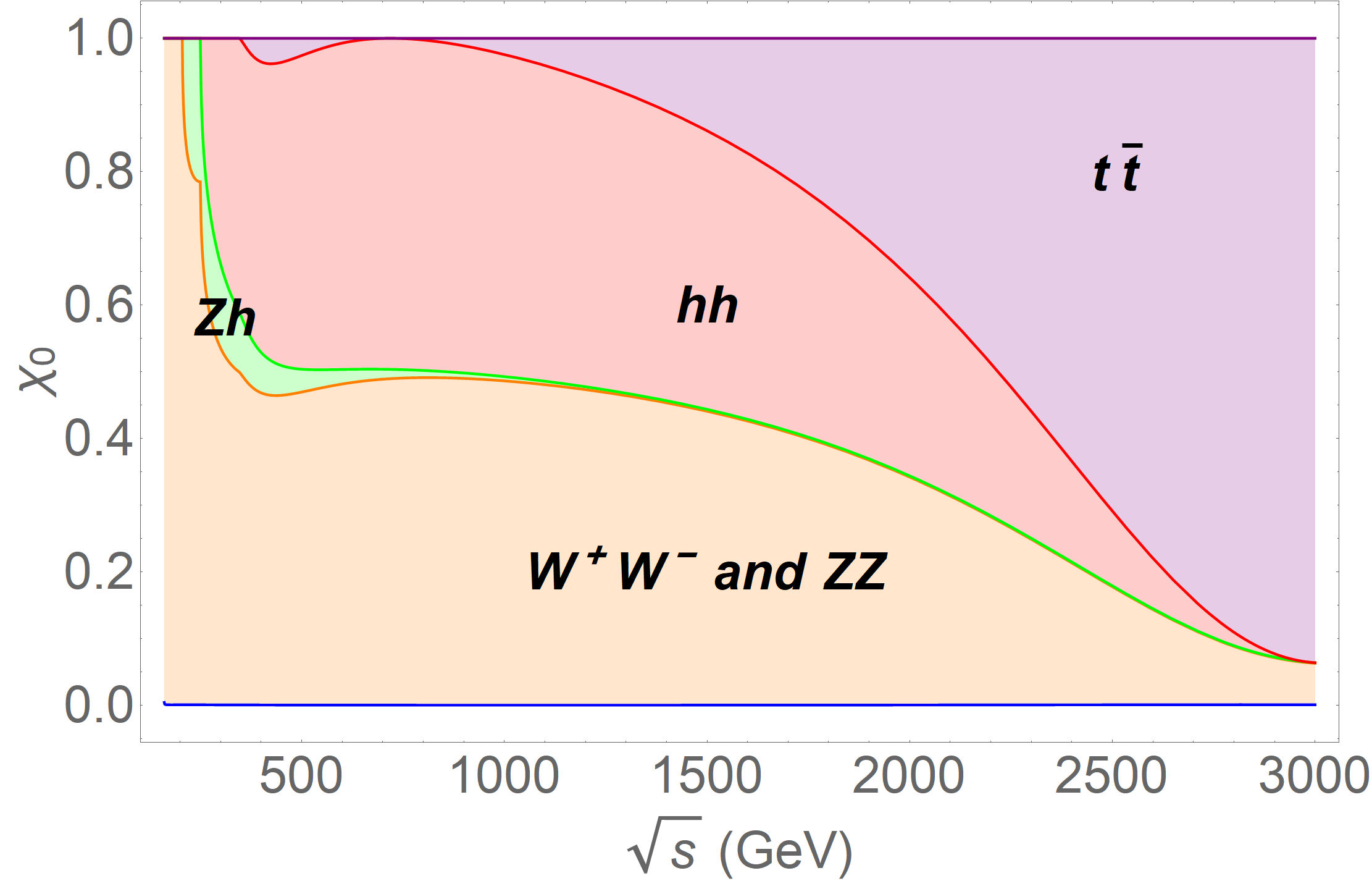}  
  \caption{$J=0$ PWA: largest fermion-loop contribution of 94\% found at 3 TeV for  $a=1.008$, $b=1.035$, $c_1=1.100$ and $d_3=0.900$.  }
  \label{fig:ratior0_bestfit_3000gev}
\end{subfigure}
\caption{}
\label{fig:ratiobestr0}
\end{figure}

\begin{figure}[!t]  
\begin{subfigure}{1\columnwidth}
  \centering
  \includegraphics[width=0.75\columnwidth]{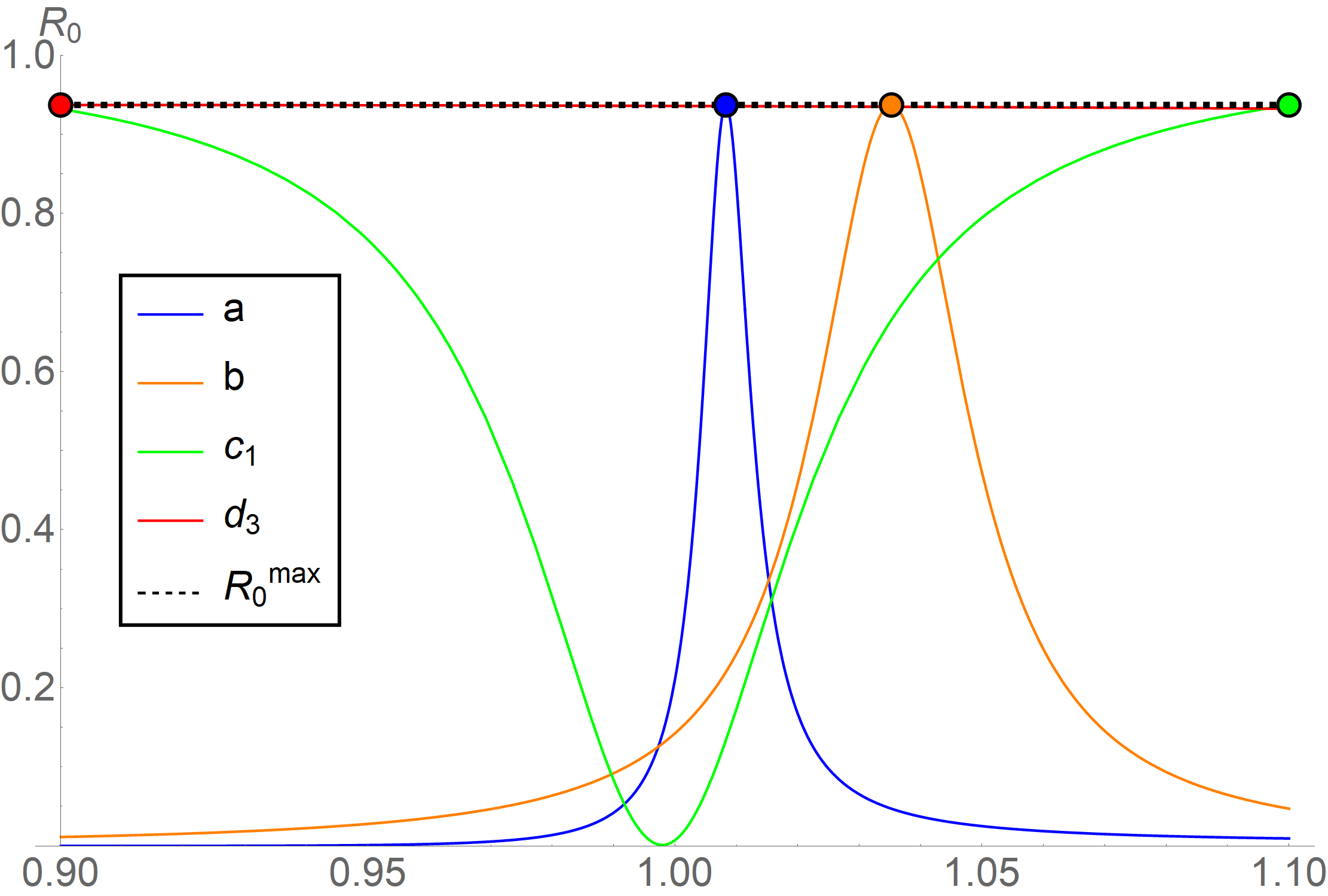}  
  \caption{}
  \label{fig:r0_best3000_sensitivity}
\end{subfigure}
\caption{\small Sensitivity of $R_0$ to each parameter when the rest are set to the  highest correction value at 
$\sqrt{s}$=3 TeV  
}
\label{fig:figr0sensitivity}
\end{figure}

If we scan the whole parameter space (10 \% deviation from the SM on the HEFT couplings) for two benchmark energies 1.5 TeV and 3 TeV,  we can find the highest fermion corrections $R_0$. 
This optimization is shown for $\sqrt{s}=3$~TeV in Fig.~\ref{fig:ratiobestr0}, where we provide the relevance of each separate cut.  
We can observe that the $b \bar{b}$ cut is negligible given the mass of the bottom quark and the structure of the interaction for the $J=0$ PWA. 
It is also important to note that 
these configurations are highly sensible to changes, as we can see for the 3 TeV benchmark energy in Fig.~\ref{fig:figr0sensitivity} . The corresponding plots for $\sqrt{s}$= 1.5 TeV show a similar behaviour.


{\bf $\bullet$ \underline{$R_1$ in the $g'=0$ limit:}}

\begin{figure}[!t]    
\centering
\begin{subfigure}{1 \columnwidth}
  \centering
  \includegraphics[width=0.8 \columnwidth]{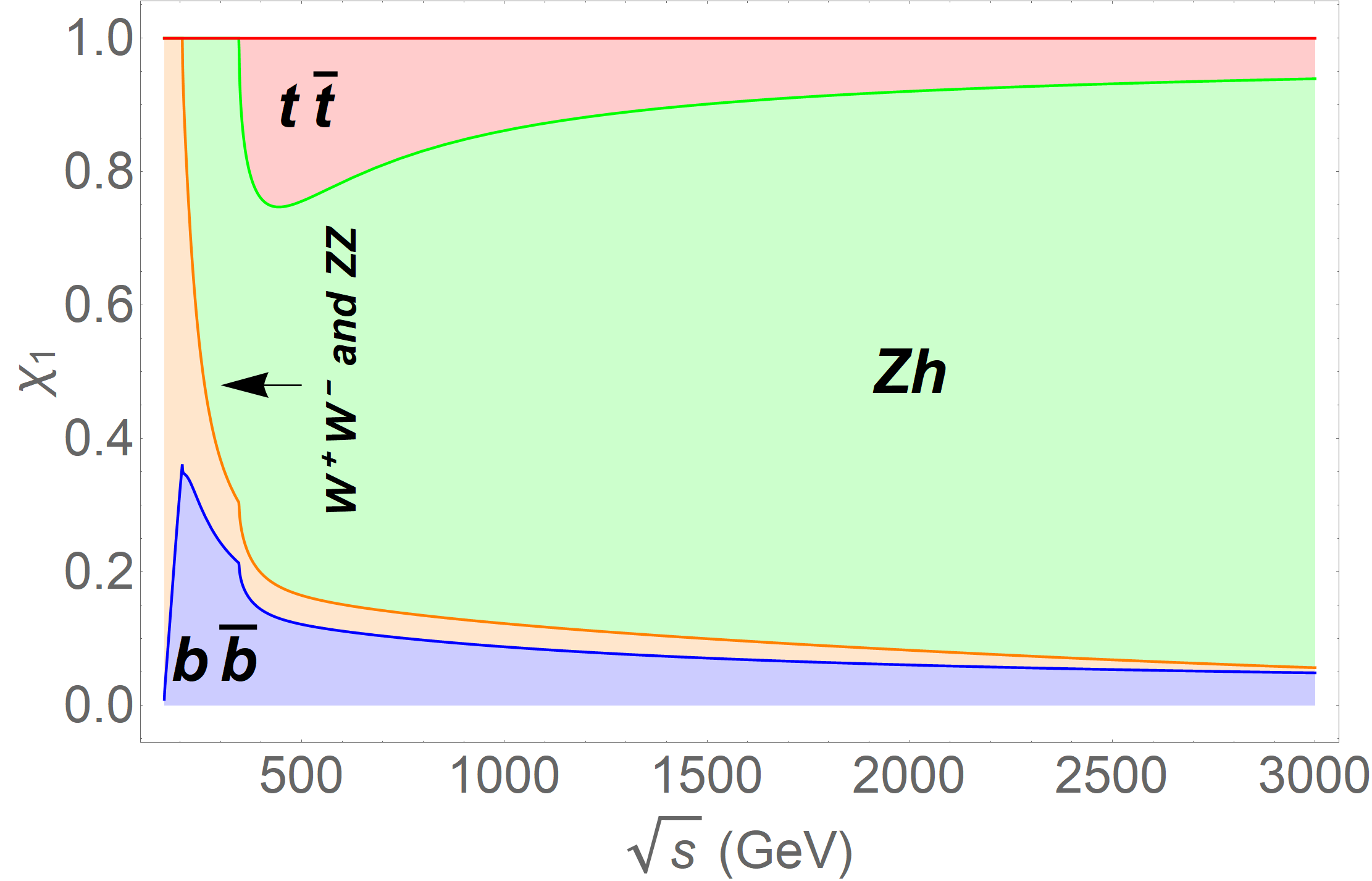}  
  \caption{$J=1$ PWA: largest fermion-loop contribution of 11\% 
  at 3 TeV for $a=1.013$ . }
  \label{fig:ratior1prime_bestfit_1500gev}
\end{subfigure}
\caption{}
\label{fig:ratior1_optimat_isospin}
\end{figure}

For this PWA we only have a dependence on $a$. Scanning for the same benchmark energies we see that the highest corrections lie close to the SM as can be seen in Fig.  \ref{fig:ratior1_optimat_isospin}. In this case the  $b\bar{b}$ cut is as relevant as the $t \bar{t}$, this means that just one of the fermions needs to be massive to produce meaningful corrections. Again, the corresponding plot for $\sqrt{s}$= 1.5 TeV show similar behavior.





{\bf $\bullet$ \underline{Beyond the $g'=0$ limit:}}

The analysis of this scenario is analogous to the previous one and we will only present the highest corrections, displayed in Tables~\ref{tab:r0summary} and~\ref{tab:r1summary}.

\vspace{-0.2cm}

\section{Conclusions}

We have calculated the top and bottom quark imaginary loop contributions to the $J=0$ and $J=1$ PWA of $W^+W-$  and compared them with the bosonic contributions to the same PWA in the context of HEFT and found that there are scenarios where these fermion contributions are meaningful and even dominant. The highest corrections are detailed in Tables \ref{tab:r0summary} and  \ref{tab:r1summary} for the mentioned PWAs. 

\begin{table}[!t]
\centering
\begin{tabularx}{1\columnwidth}{cXXXXX} 
\hline
$\sqrt{s}$  & $a-1$       & $b-1$         & $c_1-1$   & $d_3-1$     & $J=0$  \\ \hline
1.5       & 0.023  & 0.100       & \hspace{-0.1cm} -0.100    & \hspace{0.01cm} 0.100      & $R_0$=76\% 
\\ \hline
3        & 0.008  & 0.035   & \hspace{0.01cm} 0.100    & -0.100  & $R_0$=94\% 
\\ \hline
1.5     & 0.011  & 0.045    & \hspace{0.001cm}-0.100 & \hspace{0.02cm} 0.094  & $R_0'$=81\% 
\\ \hline
3        & 0.003 & 0.011 & \hspace{0.009cm} 0.100 & \hspace{0.009cm} 0.100      & $R_0'$=93\% 
\\ \hline
\end{tabularx}
\caption{\small Corrections to $J=0$ PWA for the $g'=0$ case (first two rows) and the $J=0$ p-PWA (last two rows).  In the second, third, fourth and fifth columns, we provide, respectively, the values of $a$, $b$, $c_1$ and $d_3$ that maximize the fermion-loop contributions. The center-of-mass energy is given in TeV.
}
\label{tab:r0summary}
\end{table}

\begin{table}[!t]
\centering
\begin{tabular}{ccc}    
\hline
$\sqrt{s}$   (TeV) & $a-1$       & $J=1$  \\ \hline
1.5 (PWA)      &\hspace{-0.2cm} -0.009 & $R_1$=18\%
\\ \hline
3 (PWA)         & 0.013 & $R_1$=12\%
\\ \hline
1.5 (p-PWA)       & 0.019   & $R_1'$= 66\% 
\\ \hline
3 (p-PWA)         & 0.007   & $R_1'$= 67\%   \\ \hline
\end{tabular}
\caption{\small Corrections to $J=1$ PWA for the $g'=0$ case (first two rows) and the $J=1$ p-PWA (last two rows). In the second column, we provide the value of $a$ that  maximizes $R_1$ or $R_1'$. 
}
\label{tab:r1summary}
\end{table}

By all the stated above we find that top quark and bottom loop contributions are indeed relevant when addressing higher order corrections to $W^+W^-$ scattering (and VBS in general). 
This issue should be addressed in order to have a precise description of collider experimental data.  

\end{document}